\documentclass[12pt]{article}   
\def\C{{\rm\kern.24em \vrule width.02em 
   height1.6ex depth-.05ex \kern-.26emC}}
\def\CC{{\rm\kern.1em \vrule width.02em 
   height1.07ex depth-.05ex \kern-.16emC}}
\def\R{{\rm I\kern-.20em R}}
\def\N{{\rm I\kern-.26em N}}
\def\Z{{\rm Z}\!\!\!{\rm Z}}

\newcommand{\be}{\begin{equation}}
\newcommand{\ee}{\end{equation}}
\newcommand{\bee}{\begin{eqnarray}}
\newcommand{\eee}{\end{eqnarray}}

\newcommand{\ie}{i.e.}
\newcommand{\del}{\partial}
    \newcommand{\upin}{\cup\!\! \rule{.5pt}{2.5mm}}

\def\mapdown#1{\Big\downarrow\rlap
               {$\vcenter{\hbox{$\scriptstyle#1$}}$}}

\newcommand{\xx}{x_\theta}
\newcommand{\xxx}{y_\xi}

\newcommand{\ah}{\ast_\hbar}
\newcommand{\ag}{\ast_\gamma}
\newcommand{\ce}{\check{\varepsilon}}

\newcommand{\qslc}{U_q\rm{sl}(2,\CC)}

\newcommand{\Img}{\Im_\gamma}
\newcommand{\Imm}{\Im_{\rm int}}

\newcommand{\cX}{\hat{\chi}}
\newcommand{\RRr}{\Re}
\newcommand{\RRn}{{}_N \Re}

\newcommand{\ha}{a}
\newcommand{\ad}{a^\dagger}
\newcommand{\ath}{\hat{a}_\theta}
\newcommand{\ata}{\hat{a}^\dagger_\tau}
\newcommand{\RR}{\R_\gamma}

\newcommand{\Imc}{\check{\Im}}

\newcommand{\zx}{\chi}
\newcommand{\cA}{{\cal A}}
\newcommand{\cM}{{\cal M}}
\newcommand{\cL}{{\cal L}_H}
\newcommand{\cT}{{\cal T}_2}
\newcommand{\cTT}{{\cal T}_1}
\newcommand{\cLL}{{\cal L}_H^q} 
\newcommand{\cLLd}{ {\cal L}_{H,\cdot}^q} 
\newcommand{\cLLs}{ {\cal L}_{H,\ag}^q }
\newcommand{\cLLq}{\widehat{\cal L}_q}
\newcommand{\cB}{{\cal B}}

\newcommand{\cI}{{\cal I}}
\newcommand{\qR}{{}^q\R}
\newcommand{\vR}{{}^\vee\R}
\newcommand{\cq}{\widehat{{\cal L}}_q }
\newcommand{\cAA}{\widehat{{\cal A}}  }
\newcommand{\D}{\widehat{{\cal D}}}
\newcommand{\Dq}{\cA}
\newcommand{\dd}{ {\rm d}_q  }
\newcommand{\cR}{{\cal R}}
\newcommand{\cF}{\Im}
\newcommand{\cS}{{\cal S}}
\newcommand{\zz}{\chi}
\newcommand{\sg}{\Sigma_{\check{\epsilon}}}
\newcommand{\DqN}{\cA_N(\cF:\Imm)}
\newcommand{\qRN}{\qR_N}

\makeatletter
\begin{document}
\baselineskip.7cm

\rightline{May,  2002 }
\vspace{1.0cm}
\begin{center}
\LARGE{Quantum Real Lines \\ [.15cm]
 \rule[.2cm]{1.3cm}{.7pt} Infinitesimal Structure of $\R$ 
         \rule[.2cm]{1.5cm}{.7pt}
}\\[2.3em]
\large{  Takashi Suzuki
}\\[.3cm]     
\normalsize    Hiroshima Institute of Technology            \\
\normalsize   Hiroshima 731-5143,  JAPAN\\
 \normalsize  e-mail : stakashi@cc.it-hiroshima.ac.jp

\end{center}           
\makeatother

\vspace{.15cm}

\begin{abstract}
We present in this  paper  quantum real lines as  
quantum defomations of the real numbers $\R$. 
Upon deforming  the Heisenberg algebra $\cL$ 
generated by $(a, a^\dagger)$ in terms of 
the Moyal $\ast$-product,  
we first construct $q$-deformed algebras of 
$q$-differentiable functions 
in two cases where $q$ is generic (not a root of unity) 
and $q$ is the $N$-th root of unity.  
We then investigate these algebras and finally 
propose two quantum real lines 
as the base spaces of these algebras. 
It is turned out that both quantum lines are 
discrete spaces and have noncommutative structures.  
We further find, minimal length, fuzzy structure and   
infinitesimal structure. 
\end{abstract}

\section{Introduction}

The long-standing problem, quantization of 
gravity theory, 
undoubtedly requirs some fundamental modifications 
to our understanding of geometry. 
Quantum effects must cause essential changes of 
some concepts in the classical geometry.  
Therefore, we should reconsider, for example, 
the following problems: 
(\romannumeral1) 
What is a point? 
Is it just a stable and localized object, \ie, 
of null size?  
(\romannumeral2) What is a line? 
Is it a one-dimensional object 
which is isomorphic to $\R$? 
The aim of this paper is to challenge 
these essential problems. 

As for the first problem, we expect that 
a point will get some quantum fluctuation. 
As the result, it may become something fuzzy 
and have some extra structure. 
It is natural to expect further that 
such a {\em quantum} point cannot be expressed 
just by a real number. 
As a set of  such quantum points, 
a {\em quantum} real line can be defined. 
Quantum line is, therefore, expected to 
be something fuzzy or discrete,  
and have some internal structure.  
The above observations  indicate that 
quantization of geometry cannot be 
established on a commutative ring. 
In recent developments of the  string theiry, 
nonstandard structures of geometry have appeared, 
e.g., noncommutativity of coordinates, 
fuzzy structure, and so on. 
Thus, it is quite important to make the concepts 
of geometry on a noncommutative ring clear more. 
We will refer such a geomrety as noncommutative 
geometry.

So far, a noncommutative geometry has been presented 
by Connes \cite{Co}. 
The basic strategy of the Connes' approach 
follows from the Gelfand-Naimark theorem. 
The theorem states the correspondence between a manifold $M$ 
and a {\em comutative} ring $A(M)$, 
\ie, a functional algebra on $M$. 
In precise, once a commutative algebra $A$ is given,  
a space or a manifold $M$ is always found 
and geometrical informations of $M$  can be derived from $A$. 
As an extension of this theorem,  
Connes has developped a noncommutative geometry 
according to the idea; 
if a noncommutative algebra $\cA(\cM)$ is given as a 
deformation of $A(M)$,  
the space $\cM$ which can be called 
{\em noncommutative} space will be found and 
its geometrical informations will be 
deduced from the algebra $\cA(\cM)$. 

Here, one of the manipulations to derive $\cA(\cM)$ from 
$A(M)$ is to introduce the Moyal $\ast$-product 
instead of the standard local product $\cdot$ 
into the algebra $\cA$ as the multiplication. 
On the other hand, quantum groups are deformations 
of classical algebras or groups, 
and have been expected to give some insight 
into geometry on a noncommutative ring. 
In Ref.\cite{Ma}, noncommutative geometry has been 
discussed within the framework of quantum groups. 

The author has deformed 
the classical mechanics and derived  
$q$-deformed quantum mechanics in Ref.\cite{TS1}.  
Let us review the process briefly, 
since we will follow it lattter.   
The starting point was the Poisson algebra 
$A^{\rm CM}(\Gamma)$ on the phase space $\Gamma$, 
\ie, $A^{\rm CM}(\Gamma)$ is the algebra of 
the classical mechanics. 
The algebra  $A^{\rm CM}(\Gamma)$ has been deformed 
into the algebra $\cA^{q{\rm QM}}(\Gamma_\gamma)$ of 
the $q$-deformed quantum mechanics 
where $\Gamma_\gamma$ is a deformed phase space with 
the deformation parameter $\gamma$.  
One of the key steps in the program was 
the definition of  $\Gamma_\gamma$. 
It was introduced by the product  
$\Gamma_\gamma=\Gamma\times \cT$.  
Here  $\cT$ was introduced as 
an two-dimensional spce and was      
regarded as an internal space 
attached every point on the external space $\Gamma$. 
Another important step was the definition of the 
multiplication to be introduced into 
$\cA^{q{\rm QM}}(\Gamma_\gamma)$. 
Here we have derived $\cA^{q{\rm QM}}(\Gamma_\gamma)$ 
from $A^{\rm CM}(\Gamma)$ by twice the deformations. 
The first was performed by introducing 
the Moyal product $\ah$ into $A^{\rm CM}(\Gamma)$, 
and the algebra of the quantum mechanics 
$\cA^{\rm QM}(\Gamma;\ah)$ was obtained.  
In the second step, 
$\cA^{\rm QM}(\Gamma;\ah)$ was brought 
to the final algebra  
$\cA^{q{\rm QM}}(\Gamma_\gamma;\star)$  
by the modifications $\ah\rightarrow
  \star:=\ah\otimes \ag$ 
together with 
  $\Gamma \rightarrow \Gamma_\gamma$. 
The Moyal products $\ah$ and $\ag$ act on the spaces 
of functions on $\Gamma$ and on $\cT$, respectively.  
The procedure is dipicted as 
\begin{displaymath} 
\begin{array}{ccc}
   A^{\rm CM}(\Gamma) & \buildrel 
       \ast_\hbar \over \longrightarrow 
 & \cA^{\rm QM}(\Gamma;\ah) \\ 
     \mapdown{\ast_\gamma}  &  & \mapdown{\ast_\gamma}  \\ 
 \cA^{q{\rm CM}}(\Gamma_\gamma;\ag) & 
           \buildrel \ast_\hbar \over 
         \longrightarrow & 
  \cA^{q{\rm QM}}(\Gamma_\gamma;\star)   
\end{array} 
\end{displaymath}
Notice that there exists the other path to 
$\cA^{q{\rm QM}}(\Gamma_\gamma;\star)$ via  
$\cA^{q{\rm CM}}(\Gamma_\gamma;\ag)$, 
the algebra of $q$-deformed classical mechanics.   
The essential fact is that 
the $q$-deformation comes only from 
the internal sector, \ie, 
the internal space $\cT$ and the product $\ast_\gamma$.  

In this paper, we will follow the above strategy 
to quantize the real numbers $\R$, and  
we will propose quantum real lines. 
We will start, in  Section 2, with deformation of 
the Heisenberg algebra $\cL$ 
generated by the operators $a$ and $a^\dagger$. 
The deformed algebra $\cLL$ will be proposed 
by introducing the internal sector $\cA(\cT;\ag)$  
which is the algebra of functions on $\cT$ 
and is endowed with the Moyal product $\ag$. 
We will further modify $\cLL$ to 
the operator algebra $\cLLq$ 
by changing $\cA(\cT;\ag)$ to $\widehat{\cA}(\cTT)$ 
where $\cTT$ is a reduction of $\cT$. 
In Section 3, 
we will construct $q$-deformed algebras $\Dq$ 
of $q$-differentiable functions 
from the algebra $\cLLq$.  
We will discuss $\Dq$ in two cases separately; 
the case where $q$ is generic (not a root of 1) 
in section 3.1 and 
the case where $q$ is a root of unity in Section 3.2. 
In Section 4, we will investigate the structures of 
the quantum real lines proposed as the base spaces of 
the algebras $\Dq$. 
The subsection 4.1 looks at the case when $q$ is generic 
and the quantum real line $\R_{\rm D}$ is proposed.  
We will see that $\R_{\rm D}$ is a discrete space 
and has minimal length. 
On the other hand, the quantum real line 
$\vR$ for the case with $q$ at the $N$-th root of unity is 
proposed in Section 4.2. 
Here, it will be turned out that $\vR$ consists of $\R$ 
and fuzzy internal space. 
The internal space is called the {\em infinitesimal structure}  
since it is embedded into every infinitesimal crack 
between two real numbers $x$ and $x+\epsilon\in\R$ with an 
infinitesimal $\epsilon$. 
Concluding remarkes together with future problems are 
addressed in Section 5.

\section{Deformation of the Heisenberg algebra }

Let us first devote to a deformation of 
the Heisenberg algebra $\cL$,   
which is the operator algebra generated by the two generators 
$\ha$ and $\ad$ satisfying the defining relation 
\be
\ha \ad  - \ad \ha = 1.
\label{Heisen}    
\ee
According to the strategy developped in Ref.\cite{TS1}, 
let us introduce the internal space $\cT$ and 
the algebra $\cA(\cT)$ of functions on $\cT$. 
Here $\cT$ is  a  two dimensional torus parameterized by 
$\theta$ and $\tau$. 
The generators of $\cA(\cT)$ are,  therefore, written as 
\be
U:= e^{i\theta}, \qquad V := e^{i \tau}.
\label{geneL}    
\ee
Now, we have to introduce a maltiplication 
into $\cA(\cT)$. 
The most familiar and simplest choice 
is nade by the standard pointwise product $\cdot$ 
and we denote such an algebra as $\cA(\cT;\cdot)$.     
Obviously, $\cA(\cT;\cdot)$ is 
a closed and commutative algebra, \ie, 
for ${}^\forall f(U,V),g(U,V)\in \cA(\cT;\cdot)$, 
$f \cdot g = g\cdot f \in \cA(\cT;\cdot)$. 
In this case, one identifies the base space $\cT$ with  
a {\em classical} torus.

Another possible product to be introduced in 
$\cA(\cT)$ is the, so-called, Moyal product $\ast_\gamma$ 
with some deformation parameter $\gamma$.  
Let $\cA(\cT;\ag)$ be the algebra 
endowed with the product $\ag$. 
Then, let us observe $\cA(\cT;\ast_\gamma)$ 
by giving an explicit definition of $\ag$. 
We follow Ref.\cite{TS1} where $\ag$ is defined as   
\bee
\ag & = & \sum_{n=0}^\infty  \frac{
                   \left( \frac{-i\gamma}{2}\right)^n} {n!}
\sum_{k=0}^n \left(\begin{array}{c} n \\ 
   k \end{array}\right) (-)^k 
    \overleftarrow{\partial_\theta^{\,n-k}\partial_\tau^k}
      \cdot 
    \overrightarrow
    {\partial_\theta^{k}\partial_\tau^{\,n-k}}.     
\label{gproduct}    
\eee
It is easy to see from (\ref{gproduct}) 
that the algebra  $\cA(\cT;\ag)$ is  
no longer commutative but 
a noncommutative algebra. 
Indeed, 
one obtains the basic commutation relations as 
\bee
  & U \ag V= e^{i\gamma} V \ag U&, \label{uv}   \\[.13cm]
  & \theta \ag \tau  - \tau \ag \theta =-i\gamma& . 
\label{crxiphi}    
\eee
Notice that the noncommutativity originates in 
the nonlocality of the product $\ag$, 
The commutation relations (\ref{uv},\ref{crxiphi}) 
suggests that the base space $\cT$ is 
the so-called noncommutative torus.

Now, it is the time to  construct $q$-deformed 
Heisenberg algebra $\cLL$. 
We introduce $\cLL$ by the product 
of two algebras $\cL$ and $\cA(\cT)$,   
\be
\cLL= \cL \times \cA(\cT).
\ee
In presice, let us choose the generators of $\cLL$ as  
\be
  \ath := \ha U,\qquad   
  \ata := \ad V, 
\label{defgene}   
\ee
\ie, $\cLL=\{
    \hat{\alpha} \,\vert \,
     \hat{\alpha} = \hat{\alpha}(\ath, \ata)\, \}$ 
is an operator space 
spanned by functions of operators $\ath, \,\ata$. 
We then have to define the multiplication 
to be introduced in $\cLL$. 
Upon assuming here that the sector $\cL$ possesses 
the standard operator product,  
we have only to determine the multiplication  
associated with the internal sector $\cA(\cT)$.  
In the above, we have studied two multiplications 
$\cdot$ and $\ag$. 
The product $\cdot$ makes the algebra  
$\cA(\cT;\cdot)$ commutative and $\cLL$ trivial 
in the sense that $\cA(\cT;\cdot)$ is 
factorized out from the sector $\cL$. 
Indeed, the commutation relation between 
$\ath,\,\ata \in \cLLd:=\cL\times \cA(\cT;\cdot)$ 
is the same as the relatio (\ref{Heisen}) 
in $\cL$. 
Thus, $\cA(\cT;\cdot)$ does not affect the algebra $\cLLd$ 
and $\cLLd$ is essentially equivalent to $\cL$.   
We are not interested in this case. 

On the other hand, 
we have seen that $\cA(\cT;\ag)$ 
is a noncommutative algebra.  
Owing to the noncommutativeityy, 
one expects that some nontrivial 
corrections will be made in  the operator algebra 
$\cLLs = \cL \times \cA(\cT;\ast_\gamma)$. 
Let us start the investigations of $\cLLs$ with  
the commutation relation 
between the generators $\ath,\,\ata \in \cLLs$.  
By making use of the commutation relations given in 
eqs.(\ref{Heisen}) and (\ref{uv}),  
one easily obtains the  deformed commutation relation 
\be
   \ath \ag  \ata- q\,\, \ata \ag \ath = 
                \, U\ag V, 
 \label{qcr} 
\ee
where $q$ is also the deformation parameter given by
\be
   q = e^{i\gamma}. 
\label{qvalue}    
\ee
Note here that, in the limit $\gamma \rightarrow 0$, 
the algebra $\cA(\cT;\ag)$ reduces to 
$\cA(\cT;\cdot)$ and, therefore, 
$\cLLs \rightarrow \cLLd \cong \cL$.  

Having obtained the operator algebra $\cLLs$, 
we can derive an algebra $\Dq$ 
on a $q$-deformed real numbers, \ie, $q$-deformed real line. 
Our interests are in geometrical structure of the line. 
According to the strategy of the noncommutative 
geometry \cite{Co}, 
the structure will be extracted from $\Dq$. 
These investigations are the tasks in the following sections. 
However, for the latter discussions, 
it is convenient  to modify $\cLLs$ 
to the new algebra $\cLLq$ 
in which the internal sector 
is {\em not} the functuibal algebra $\cA(\cT;\ag)$ 
{\em but} an operator algebra $\cAA(\cTT)$. 
The base space of $\cAA(\cTT)$ is 
a one-dimensional space $\cTT$. 
We will define $\cAA(\cTT)$ so that we have a similar 
commutation relation to (\ref{uv}). 
Furthermore, upon defining the algebra $\cLLq$ by 
\be 
  \cLLq := \cL \times \cAA(\cTT),  
\label{defcTT}    
\ee
we require that the defining relation is  
similar to (\ref{qcr}). 
Before going to the explicit definition of $\cAA(\cTT)$, 
we should give a comment on the essence  
of the modification from $\cLLs$ to $\cLLq$.  
We have seen that, in the algebra $\cLLs$, 
the $q$-deformation originates in the 
nonlocal structure of the product $\ag$.  
On the contrary, we require that the same 
$q$-deformed effects come from the noncommutativities 
originating in the operator nature of $\cAA(\cTT)$. 
Here, we assume that the standard operator product 
is introduced in $\cAA(\cTT)$. 
In presice, upon denoting the generators of $\cAA(\cTT)$ as 
$\hat{U}$ and $\hat{V}$, 
the basic commutation relation 
\be
\hat{U} \hat{V} = q \hat{V} \hat{U}
\label{quv}   
\ee
instead of eq.(\ref{uv}) yields $q$-deformed 
effects in $\cLLq$. 

Let us go ahead to the explicit representations of 
$\cAA(\cTT)$ and $\cLLq$. 
By writing the generators of $\cAA(\cT)$ as 
$\hat{U}=e^{i\hat{\theta}},\, \hat{V}=e^{i\hat{\tau}}$, 
the required relation (\ref{quv}) is satisfied if 
\be  
\hat{\theta} = \theta, \qquad 
\hat{\tau}= i\gamma \frac{d}{d\theta} := 
    i\gamma \del_\theta.   
\label{phi}     
\ee 
Therefore, the base space $\cTT$ of the algebra $\cAA(\cTT)$ 
is selectted as a one-dimensional torus 
parameterized by $\theta$.  
Notice that the representations given in (\ref{phi}) 
are consistent with the commutation relation 
corresponding to (\ref{crxiphi}),  
\ie,  
$\hat{\theta}\hat{\tau} - \hat{\tau}\hat{\theta} 
   = -i\gamma$.  
Next, let $\cLLq$ be generated by the operators 
$\ha_q, \ad_q$ which  are defined 
by the substitutions  
\be
  \ha_\theta\,\longmapsto\,\ha_q, \qquad 
  e^{i\theta}\,\ad_\tau\,\longmapsto\, \ad_q.  
\label{changes}   
\ee
The commutation relation between these generators 
is finally obtained as
\be
\ha_q \ad_q -q\, \ad_q \ha_q = q^{i\del_\theta}.
\label{newcr} 
\ee
We have reached the final $q$-deformed Heisenberg algebra 
$\cLLq$ 
with the deformation parameter $q$ given in (\ref{qvalue}). 
It is important to notice that 
the variable $\theta$ becomes the coordinate 
of the internal space of the Hilbert space  
on which $\cLLq$ acts.  

In the next section, 
we will realize $\cq$ explicitly by $\D$,  
the algebra generated by 
a coordinate operator and a differential operator 
on $q$-deformed real numbers. 
We then represent $\D$ by 
the algebra $\Dq$ corresponding to the   
Hilbert space of $\D$, \ie, 
$\Dq$ is the algebra of $q$-differentiable functions.

\section{Deformations of algebra of functions on $\R$}

Recall that a posible 
representation of the Heisenberg algebra $\cL$ 
is realized by the replacements $\ha\rightarrow \del_x$ and 
$\ad \rightarrow x\in \R$. 
We will adopt similar replacements to the $q$-deformed 
algebra $\cLLq$. 
Let $\D$ be the operator algebra generated by $q$-deformed 
coordinate and $q$-deformed differential operators. 
The algebra $\D$ can be derived from $\cq$ by the replacements   
\be
{}\hskip-.5cm
\ha_q \,\longrightarrow \, D_q, \qquad 
\ad_q\, \longrightarrow \, \cX
\label{replace}    
\ee
where $D_q$ and $\cX$ stand, respectively, for 
the $q$-differential and the $q$-cordinate operators 
on a quantum real line.  
Our goal is to reveal geometry of the line. 
In order to reach there, 
we have to study the $q$-deformed 
algebra $\Dq$ on which the action of $\D$ is defined. 
In the following subsections, we are going to discuss 
$\Dq$ for two cases, separately; 
in section 3.1, the case where $q$ is  generic and 
in section 3.2, the case with  $q$ at a root of unity.

\subsection{The case  where $q$ is generic}

Let us suppose first that the parameter $q$ 
is not a root of unity. 
Upon the substitutions (\ref{replace}), 
the commutation relation (\ref{newcr}) reads 
the commutation relation betwen $D_q$ and $\cX$ as
\be
D_q \cX - q\, \cX D_q = q^{i\del_\theta}.
\label{com}     
\ee
and the explicit expressions of these operators 
are presented by 
\be
\cX =  \frac{1+ q^{i\del_\theta}}{1+q^{-1}} \, 
                               x_\theta, \qquad 
D_q = \frac{q^{-i\del_\theta}-1}
                           {x_\theta(q-1)}. 
\label{defaa}   
\ee
Here, we temporarily give the base space of these operators 
as
\be
  \RRr=\{ \xx \,\vert \,
     \xx := x_+ e^{i\theta}, \, \, 0<x_+ \in\R,\,  
    0\leq \theta < 2\pi \}. 
\label{qcoordinate}     
\ee
One then construct a functional algebra  $\Im^q$ 
on $\RRr$  as 
\be
\Im^q=\{ f(\xx)\,\, \vert \,\, \xx\in \RRr,
     \},  
\ee 
where the multiplication is supposed to be the standard 
pointwise product $\cdot$. 
The complete basis $\cB$ of $\Im^q$ is given by
\be
\cB=\{\xx^n\,\,\vert \,\, n\in \Z\} 
\label{basis}    
\ee
and, therefore, ${}^\forall f(\xx) \in \Im^q$ is expanded as 
\be 
f(\xx) = \sum_{n\in \Z} f_n\, \xx^n .
\ee
The actions of the operators $\cX$ and $D_q$ on $\Im^q$  
are calculated explicitly by (\ref{defaa}), \ie,     
\be
\cX\, f(\xx) = \xx \, 
     \frac{f(\xx) + q^{-1}f(\xx q^{-1})}{1+q^{-1}}, \qquad
D_q f(\xx) =
           \frac{f(\xx q) -f(\xx)}{\xx(q-1)}.
\label{defxiD}    
\ee
Note that, by taking  the limit $q\rightarrow 1$ 
together with $\theta \rightarrow 0$, 
$\cX$ becomes the standard coordinate operator 
$\hat{x}$ such as $\hat{x}f(x) = x f(x)$,   
and $D_q$ reduces to the standard differential operator, \ie,  
$D_qf(\xx) \stackrel{q\rightarrow}{\longrightarrow} 
     \frac{d}{dx}f(x)$. 
Furthermore, one shows that 
the $q$-differential operator $D_q$ 
satisfies the following properties,   
\bee
\mbox{}\hskip-1cm
&& D_q \, {\bf 1} =0, 
   \quad  \mbox{for the identity}\,\,{\bf 1}\in\Im^q,  
             \label{constant}            \\[.15cm]
 && 
    D_q \left( f\cdot g \right) = 
     \left( D_q f(\xx) \right) g(\xx) 
     + f(q\xx) \left( D_q g(\xx) \right). 
  \label{Leibniz}    
\eee 
The first equation (\ref{constant}) indicates that 
$D_q$ vanishes constants in $\Im^q$, 
and
 (\ref{Leibniz}) shows 
the deformed Leibniz rule. 

Now, one gives the $q$-deformed algebra $\Dq(\Im^q;D_q)$ 
as the algebra of functions 
on which the action of the operator $D_q$ is defined, \ie,  
$\Dq(\Im^q;D_q)$ is the algebra of $q$-differentiable functions 
on $\RRr$. 
The base space $\RRr$ is the object which is called 
\lq\lq quantum real line" and 
its geometrical structure is derived from $\Dq(\Im^q;D_q)$. 
By the definition given in (\ref{qcoordinate}), 
$\RRr$ looks like the complex plane $\C$. 
However, we should notice that  the algebra $\Dq(\Im^q;D_q)$ 
possesses only the {\em difference} operator. 
Therefore, $\RRr$ may not be the standard complex plane but 
have some discrete structure. 
It will be turned out, in Section 4, 
that the final quantum real line,  
which will be derived  from $\RRr$ by some modifications,    
has such a discrete structure.

Let us end this subsection with the study of 
$q$-differential structue in $\Dq(\Im^q;D_q)$. 
Notice first that, as far as $\Im^q$ is considered, 
the internal sector does not play any special role. 
Indeed, for ${}^\forall f(\xx), g(\xx)\in \Im^q$, 
$f\cdot g= g\cdot f$,  \ie, $\Im^q$ is still a 
commutative algebra.  
However, once the operator $D_q$ is taken into accout, 
\ie,  the algebra $\Dq(\Im^q;D_q)$ is considered, 
some nontrivial structures appear as shown above. 
It is interesting to show further that 
the exterior derivative $\dd$ can be introduced 
in $\Dq(\Im^q:D_q)$ by
\be
\dd f  = \dd \xx D_qf(\xx).  
\ee   
By operating $\dd$ on the identity 
$\xx\cdot f(\xx) = f(\xx)\cdot \xx$, 
another noncommutativity 
\be
\dd \xx\,f(\xx) = f(q\xx)\,\dd \xx
\label{NC}   
\ee 
is found. 
Thus, one regards the algebra $\Dq(\Im^q;D_q)$ 
as a noncommutative algebra.

\subsection{The case where $q$ is a root of unity}

Let us next turn to the $q$-deformed  algebra $\Dq$ 
in the case where $q$ is given by the $N$-th root of unity, \ie, 
\be
       q = e^{i\,\frac{2\pi}{N}}.
\label{rootofunity}     
\ee
Let $\Dq_N(\Im^q_N;D_q)$ be the algebra of 
$q$-differentiable functions to be discussed in this case.  
Here $\Im^q_N$ represents the functional algebra on 
the base space $\RRn$ given below. 
In the case with (\ref{rootofunity}),  
it is natural to expect that 
the base space is different from $\RRr$ 
for the case with generuc $q$.  
Now, let us denote the base space as 
\be
\RRn =\{\, \xxx\, \vert \, \xxx := y\xi, 
      \,\,\, \xi=e^{i\theta}  \},   
\label{qqline}     
\ee
\ie, the algebra $\Im_N^q$ on $\RRn$ is given by 
\be
   \Im^q_N = \{ 
   f(\xxx) \,\, \vert \,\, 
    f(\xxx) = \sum_{n\in \Z} \alpha_n\, \xxx^n,  \,\, \,
    \xxx\in\RRn,  
  \},    
\label{defImqN}     
\ee
with the basis 
\be
\cB_N:=\{\,\xxx^n\,\vert\, n\in \Z\,\}. 
\label{baseN}   
\ee

In order to make $\Im_N^q$ and $\Dq_N(\Im_N^q;D_q)$ explicit,  
we have to observe first the operator algebta 
$\D_N$ which acts on $\Dq_N(\Im_N^q;D_q)$. 
We expect that $\D_N$ is drastically different from 
$\D$ in the case with generic $q$. 
Indeed, it is well-known that 
the representations of quantum groups or quantum universal 
envelopping algebras with $q$ at a root of unity 
are completely differet from those with generic $q$. 
For example, in a quantum representation 
$\cR(G)$ of the group $G$, 
a classical (undeformed) sector $R(G)$ appears    
apart from a $q$-deformed sector $\cR'(G)$ 
such as $\cR(G)=R(G)\otimes\cR'(G)$. 
We will see later that such a remarkable structure  
appear in our algebras $\D_N$ and $\Dq_N(\Im^q_N;D_q)$. 

As in the previous case in Section 3.1, 
the $q$-differential operator $D_q$ is also 
introduced by the second equation in (\ref{defxiD}) 
with the replacement of the variable 
$\xx \rightarrow \xxx$. 
However, characteristic features of the case 
with $q$ at a root of unity appear  
through the following equations;  
\bee
D_q \xxx^{kN+r} &=& [r]\,\xxx^{kN+r-1}, \quad \mbox{for} 
      \quad r \neq 0, \nonumber \\[.15cm]
 D_q \xxx^{kN} &=& 0.
\label{operationD}  
\eee
One sees from (\ref{operationD}) that $\xxx^N$ is 
a constant with respect to $D_q$. 
Further, one finds easily that the action of $D_q^N$ on 
${}^\forall f(\xxx)\in \Im^q_N$ is null.  
However, upon defining another operator $\del$ by 
$\del := D_q^N/[N]!$, 
the actions of $\del$ on $\cB_N$ are calculated as  
\bee
\del\, \xxx^{kN;r} = k\,\xxx^{(k-1)N+r}. \qquad 
      && \mbox{for} \quad  k\geq 1 \nonumber \\[.15cm]
\del\, \xxx^r =0, \qquad \qquad \qquad 
      && \mbox{for} \quad  r\leq N-1. 
\label{Naction}    
\eee
where use has been made of the relation $[kN]/[N]=k$. 
Notice further that $\xxx^r,  \, r\leq N-1$ 
behave as constants with respect to the operator $\del$. 
It should be emphasized that,  
as can be seen from (\ref{operationD})  
and (\ref{Naction}), 
these two operators $D_q$ and $\del$ are 
independent of each other on $\Im^q_N$.   
Therefore, in order to define the operator algebra $\D_N$, 
$\del$ is needed as well as the $q$-differential 
operator $D_q$. 

Let us consider  the algebra $\Im^q_N$ 
on which $\D_N$ acts. 
The above observations indicate that 
there exists the mapping $\pi$ such as
\be
 \begin{array}{cccl}
 \pi\,\,: &  \Im^q_N  &  \longrightarrow  & \cF  \otimes \Imm
                                        \\[-0.2cm]
   & \upin  &         & \mbox{}\quad {\upin}
                                       \\[-.2cm]
  & f(\xxx)  & \longmapsto & \pi(f) = \lambda(x)  \psi(\xi).
  \end{array}
\ee
Here the variable $x$ is introduced  by  
\be
\pi(\xxx^{kN+r}) = x^k\xi^r 
\ee
with $k\in\Z$ and $r$ being valued in  
\be
   r \in {\cal I} := \left\{ 
  \begin{array}{ll} 
      \{ -p , -p+1, \cdots , p-1, p \},
     & \mbox{for} \,\,\, N=2p+1 , \\[.2cm] 
      \{-p+1, \cdots, p-1, p \},    & \mbox{for} \,\,\, N=2p
      \end{array}  \right. .
\label{region}     
\ee
The basis $\cB_N$ is, therefore, factorized into two sets as 
$\cB_N \rightarrow {\cal W} \otimes \varpi$ where 
${\cal W} := \{
   x^k\,\vert\, k\in \Z\}$ and  
$ \varpi := \{  \xi^r\,\vert\, r \in {\cal I} \}$.  
Namely, any functions $\lambda(x) \in \cF$ 
and $\psi(\xi)\in \Imm$ are expanded as   
\be
  \lambda(x) = \sum_{k\in \Z} \lambda_k\, x^k, \qquad  
  \psi(\xi) = \sum_{r\in {\cal I}} \psi_n \xi^r. 
                 \label{expansion1} 
\ee   
Now, we have reached the important fact that  
the base space of $\Im_N^q$ should be 
introduced instead of $\RRn$ as 
\be
\qR \ni (x,\xi), 
\label{qR}   
\ee
whose gometrical structure will be clarified 
in the next section. 

Let us go back again to $\D_N$ and 
observe how the operators in $\D_N$ acts on 
the factorized space $\cF \otimes \Imm$.     
The mapping $\pi$ induces anothe mapping 
$\hat{\pi}\,: \,  \D_N \rightarrow 
      \D_N^{cl} \otimes \D_N^{\rm int} $. 
The equations in  (\ref{Naction}) show that the operator 
$\del$ acts on $\cF\otimes\Imm$ as  
$\hat{\pi}(\del) =\del_x\otimes{\bf 1}$, 
$\del_x$ being the standard differential operator,   
while from eq.(\ref{operationD}), one sees that  
$\hat{\pi}(D_q)={\bf 1}\otimes D_q$, \ie,  
\be
  \del_x\lambda (x) = \frac{d}{dx} \, \lambda(x), \qquad 
  D_q\,\psi(\xi) = \frac{\psi(q\xi)-\psi(\xi)}{\xi(q-1)}.
\label{dif}     
\ee

We are now at the stage to discuss the $q$-deformed  
algebra $\Dq_N(\Im_N^q;D_q)$. 
The investigations given above indicate that 
$\Dq_N(\Im_N^q;D_q)$ is essentially factorized  as 
\be
\Dq_N(\Im^q_N;D_q) \cong \DqN:=
A(\cF; \del_x) \otimes 
                    \cA_{{\rm int}}(\Imm;D_q).  
\label{algebraA}             
\ee
The first sector $A(\cF;\del_x)$ can be regarded as 
the external sector and as the standard 
(undeformed) algebra of differentiable functions,   
where the variable $x$ is real number.  
Namely, the base space of $A(\cF;\del_x)$ is just  $\R$.  
On the other hand, $\cA_{{\rm int}}(\Imm;D_q)$ is regarded   
as the internal sector 
and is an algebra of $q$-differentiable functions. 
The base space of $\cA_{{\rm int}}(\Imm;D_q)$ is 
expected to have some nonstandard structure 
on which we will discuss later.  
Thus, in the case where $q$ is a root of unity, 
the classical sector appears explicitly 
appart from the $q$-deformed internal sector. 

Finally, we should introduce a suitable 
multiplication in $\cA_{{\rm int}}$. 
Notice here that the algebra $\cA_{{\rm int}}(\Imm;D_q)$ 
is not closed with respect to the standard 
pointwise product  $\,\cdot\,$. 
Indeed, for $r,\,r' \in \cI$, the value $s$ such as 
$\xi^r \cdot \xi^{r'}=\xi^s$ is not always in $\cI$. 
In order to make ${\cA}_{{\rm int}}(\Imm;D_q)$ 
a closed algebra, let us introduce the mapping 
$\mu : \cA_{\rm int}\,\rightarrow\, 
       \hat{\cA}(M)$. 
Here $\hat{\cA}(M)$ is the algebra of 
$N\times N$ matrices $M$ generated by  
the basis   
\be 
     \hat{\xi}^r = \mu(\xi^r) = 
    \left(
      \begin{array}{cc}
                 &   I_{N-r}  \\
             I_r &    
     \end{array}
   \right),  
   \quad r\in {\cal I}, 
\label{defop}     
\ee
where $I_n$ is the $n\times n$ unit matrix 
and the convension $I_{N+n}=I_n$ for $n>0$ is used. 
Notice that the algebra $\hat{\cA}(M)$ 
is closed under the standard matrix multiplication, 
and $\hat{\varpi}=\{\hat{\xi}^r \,\vert\, r\in \cI\}$ 
is a complete basis. 
Further, $\hat{\varpi}$ is  
orthonormal with respect to the pairing 
$\langle A,B\rangle := A^\dagger B$. 
Hrere we use  $(\hat{\xi}^r)^\dagger= \hat{\xi}^{-r}$ 
but $(\hat{\xi}^p)^\dagger =\hat{\xi}^p$ 
only for  the case $N=2p$.  
Since the mapping $\mu$ is an isomorphism 
between $\cA_{\rm int}(\Imm)$ and $\hat{\cA}(M)$,   
the multiplication $\cA_{\rm int}(\Imm)$ should 
possess is introduced as follows; 
upon denoting the multipliction as $\odot$, 
the product of two functions 
$\psi, \, \psi'\in \cA_{\rm int}(\Imm;D_q)$ is defind by  
\be
\psi(\xi) \odot \psi'(\xi) := \mu^{-1}\left( \mu(\psi)\mu(\psi')
    \right)   
\label{multi}     
\ee
Now we have understood the $q$-deformed  
algebra $\DqN$ on $\qR$. 
The geometrical structure of the base space $\qR$ 
will be discussed in the next section.

\section{Quantum real lines}

We have obtained the algebras of $q$-differentiable functions;  
for the case with generic $q$, we have $\Dq(\Im^q;D_q)$  
on the base space $\RRr$, 
and for the case with $q$ at the $N$-th root of unity, 
we have $\DqN$ on  $\qR$.  
Let us go ahead to the main task of our progtam, 
\ie, the investigations of geometrical structures of 
these base spaces. 
This is performed via the algebras 
$\Dq(\Im^q)$ and $\DqN$.  
We will finally propose quantum real lines 
$\R_{\rm D}$ when $q$ is generic and 
$\vR$ when $q$ is a root of unity.  

\subsection{Quantum real line $\R_{\rm D}$ ;  
   $q$ is not a root of unity }

Notice first that, as we have seen in Section 3.1,  
the algebra $\Dq(\Im^q;D_q)$ possesses the difference structure 
induced by $D_q$. 
We can, therefore, regard the base space $\RRr$ 
as a discrete space. 
To see this in more explicit,  
let us chage the variable from $\xx$ to $\zz$ 
by the relation 
\be
\zz= -i \,\log \xx. 
\label{defzx}   
\ee 
Now, let us denpte the base space parameterized by 
$\zz$ as $\RR$. 
According to the change of variable, 
the difference operator $D_q$ with respect to $\xx$ 
should  be changed to the operator $\Delta_\gamma$ 
with respect to $\zz$. 
One assumes the relation 
$D_q=(D_q\zz) \Delta_\gamma$ and finds 
\be 
\Delta_\gamma= \frac{e^{\gamma\,\zz\del_{\zz}}-1}{\gamma}. 
   \label{difgamma}   
\ee
Regarding $\Im^q$ as the space of 
functions on $\RR$, 
the action of $\Delta_\gamma$ on a functions 
$f(\zz) \in \Im^q$ is explicitly written as 
\be
\Delta_\gamma  f(\zz) = 
     \frac{ f(\zz+\gamma) - f(\zz) }{ \gamma }.
\ee
The key commution relation between the coordinate $\zx$ 
and $\Delta_\gamma$ is calculated as  
\be
\chi \Delta_\gamma - \Delta_\gamma \chi = {\cal K},  
\ee
where ${\cal K}$ is the shift operator such as 
${\cal K}f(\zx)=f(\zx+\gamma)$. 

On the base space $\RR$, 
we have the algebra $\Dq(\Im^q;\Delta_\gamma)$ as  
the space of functions on which 
$\Delta_\gamma$-action is defined.  
It should be emphasized that the variable $\zz$ 
is, at this stage, thought to be  continuous. 
However, upon choosing a value $\zz_0$ arbitrarily, 
one can reduce $\Im^q$ to the subset 
$\Im_{\rm D}(\zz_0)=\{ f(\zz_0+n\gamma)\,\vert \,  n\in \Z\}$ 
owing to the difference operator $\Delta_\gamma$. 
Namely, we can restrict ourselves to the subalgebra 
$\Dq(\Im_{\rm D}(\zz_0);\Delta_\gamma) 
     \subset \Dq(\Im^q;\Delta_\gamma)$.   
According to the restriction of the algebra, 
the base space is also restricted to a discrete space 
with equal distance $\gamma$. 
The base space 
of $\Dq(\Im_{\rm D}(\zz_0);\Delta_\gamma)$ 
is, therefore,  written as  
$\R_{\rm D}(\zz_0)=\{\zz_0+n\gamma \,\vert\, n\in\Z\}. $ 
If one choose another point $\zz_0' \in \RR$, 
then the algebra $\Dq(\Im_{\rm D}(\zz_0');\Delta_\gamma)$ is 
obtained as another subalgebra of 
$\Dq(\Im^q;\Delta_\gamma)$ and  the base space  
$\R_{\rm D}(\zz_0')$ is derived.  
As a matter of course, 
the algebras 
$\Dq(\Im_{\rm D}(\zz_0);\Delta_\gamma)$ and 
$\Dq(\Im_{\rm D}(\zz'_0);\Delta_\gamma)$ are equivalent.  
Dividing $\Dq(\Im^q;\Delta_\gamma)$ 
by the equivalence,   
one has finally reached the following proposition: \\[.3cm]
{\bf Proposition 1} : Quantum Real Line $\R_{\rm D}$ 
                                                    \\[.15cm] 
{\em In the case where the deformation parameter $q$ is not 
a root of unity, 
the quantum real line $\R_{\rm D}$ is a discrete space 
composed of an infinite number of points $\zz_n$ such as} 
\be
\R_{\rm D} := \{ \zx_n\, \vert \,  \zx_n =n\gamma +\zx_0,  
  \,\,  \,\,\,n \in \Z \,\}.
\ee 
\rule{13.55cm}{0pt} \rule{2mm}{2mm} \\[.13cm]
The quantum real line $\R_{\rm D}$ is dipicted in Fig.1. 
\begin{displaymath}
\rule[.1cm]{10cm}{.38pt} \hskip-.365cm>\,\,\,\R_{\rm D}
  \hskip-9.8cm 
  \begin{array}{ccccccc}
     & & & & & \\[-.1cm]
    & \bullet 
    & \bullet 
    & \bullet 
    & \bullet 
    & \bullet 
    &                           \\
  \cdots & \zx_{n-2}  & \zx_{n-1} & \zx_n  
        & \zx_{n+1}  & \zx_{n+2} & \cdots
   \end{array} 
   \rule{3cm}{0pt}
\end{displaymath}

\vspace{.15cm}
\begin{center}
Fig.1  The quantum real line $\R_{\rm D}$
\end{center}
%

Let us look at the situations in the limit 
$\gamma \rightarrow 0 \,(q \rightarrow 1)$.  
We find first that the intervals between adjacent two points 
become zero and the quantum line $\R_{\rm D}$ 
becomes a continuous line. 
Further, the difference operator given in (\ref{difgamma}) 
reduces to the standard  differential operator $\del_{\zz}$ 
with respect to the variable $\zx$. 
From these observations, one can conclude that,  
in the limit $\gamma \rightarrow 0$, 
the variable $\zx$ corresponds to the standard real number 
and, therefore, $\R_{\rm D} \rightarrow \R $. 

It is important to discuss the relation 
between $\RR$ and $\R_{\rm D}$. 
As we have seen, 
the algebra $\Dq(\Im^q;\Delta_\gamma)$ on $\RR$ is 
the direct sum of an infinite number of subalgebras 
$\Dq(\Im_{\rm D}(\zx_0);\Delta_\gamma)$ 
with $0\leq\zx_0<\gamma$.  
This fact immediately  derives that   
$\RR$ is also the union of $\R_{\rm D}(\zx_0)$, 
\ie, $\RR=\cup_{0\leq\zx_0<\gamma} \R_{\rm D}(\zx_0)$. 
One can say from this relation between $\RR$ and 
$\R_{\rm D}$ as follows: 
A quantum real line $\R_{\rm D}$ is nothing but 
a representative, 
In this sense, each point $\zx_n\in\R_{\rm D}$ is 
a representative, 
\ie, an infinite number of $\zx_n$ fill 
the interval, denoted as ${\rm I}_n=[n.n+1)\subset \RR$,  
of width $\gamma$. 
One cannot distinguish a point $\zx_n\in {\rm I}_n$ 
from others.  
In other words, the location of the point $\zx_n$ 
is  always specified with the ambiguity of width $\gamma$. 
This suggests that we always have the uncertainty 
$\Delta\chi \sim\gamma$ in coordinate of the quantum line, 
\ie, there appears  {\em minimal length} $\gamma$.

As the final discussion, 
let us observe $\RR$ and $\R_{\rm D}$ from another viewpoint. 
One should notice that the space $\RR$ 
parameterized by the continuous variable $\zz$ 
can be regarded as a cylinder $\R \times S^1$. 
Each point in $\R$, \ie, each real number has 
an internal space $S^1$. 
Conversely, each point in $S^1$ has a real line $\R$.  
Namely, an infinite number of real lines attach 
to $S^1$.     
One can freely pick up just one real line $\R$ among them.  
The quantum real line $\R_{\rm D}$ is identified with 
the real line $\R$ we have picked up 
and an infinite number of $S^1$ on the $\R$. 
The interval ${\rm I}_n$ corresponds to $S^1$, 
\ie, the size of the internal space $S^1$ 
is $\sim\gamma$. 
Taking the minimal length  discussed above into account, 
one may consider that quantum point looks 
a one-dimentional circle.

\subsection{Quantum real line  $\vR$ ; 
      $q$ is a root of unity}

Let us turn our attention to the case where 
the deformation parameter $q$ is the $N$-th root of unity. 
In Section 3.2, we have obtained the closed algebra $\DqN$ and 
introduced the base space 
$\qR= \{ (x,\xi)\, \vert \, x\in \R, \,\, 
\xi=e^{i\theta}, \,0\leq \theta < 2\pi\,\}$ in (\ref{qR}). 
We are at the stage to investigate geometrical  structure 
of the space $\qR$ 
and to propose the final quantum real line $\vR$.  
Since $\DqN$ is the tensor product of two spaces 
$A(\cF;\del_x)$ and $\cA_{\rm int}(\Imm;D_q)$, 
the base space $\qR$ is expected to be  the tensor product of 
$\R$ and $\cS$ which is introduced as 
the base space of $\cA_{\rm int}$. 
What we should do first is to study the geometrical 
structure of the internal spaca $\cS$.  

To show the structure of $\cS$, 
one should notice that the basis 
$\varpi$ 
of $\cA_{\rm int}$ is of finite dimension, 
\ie, ${\rm dim.}\hat{\varpi}=N$, and 
recall that $\varpi$ is complete.  
Therefore, any function in $\cA_{\rm int}$ can be 
expanded uniquely in terms of $\xi^r, \, r\in \cI$. 
In particular, there exists the function 
$\delta(\theta;\theta')\in \cA_{\rm int}$ 
which plays the role of the \lq\lq delta function" as
\be
\delta(\theta; \theta') := 
\sum_{r\in {\cal I}} \left( \xi^{\ast} \right)^r \cdot
      \left( \xi' \right)^r = 
  \sum_{ r\in {\cal I}} e^{ir(\theta'-\theta)} .
\label{delta}     
\ee
Actually, in the limit of $N\rightarrow \infty$, 
the function $\delta(\theta ; \theta')$ returns to the 
Dirac's delta function as
\be
 \lim_{N\rightarrow\infty} \delta(\theta;\theta') = 
     \delta(\theta-\theta'). 
\ee
This suggests that, for a  point $\theta=\theta'$,   
the function $\delta(\theta;\theta')$ 
does not have a sharp peak 
but spreads around the point. 
In this sense, the base space $\cS$ is composed of 
\lq\lq fuzzy" points 
and is regarded as a \lq\lq fuzzy" circle. 
Thus, the internal space of $\qR$ is the fuzzy circle 
and $\qR$ is written as, 
\be 
         \qR= \R \times \cS, 
\label{qRfinal}        
\ee
\ie, every point $x \in \R$ has the fuzzy internal space $\cS$. 

Having obtained the base space $\qR$ as  (\ref{qRfinal}),  
we should make clear the relationship 
between the internal space $\cS$ 
and the external space $\R$. 
Recall the following facts: 
The difference operator $D_q$ acting on 
the internal sector $\Imm$ 
generates the finite displacement $\xi \rightarrow q\xi$. 
By the definition 
$q=e^{2\pi i/N}$, one finds 
$D_q : (x,\theta) \hookrightarrow (x,\theta +\frac{2\pi}{N})$, 
where ${\rm P} \hookrightarrow {\rm P}'$ 
stamds for the mapping 
from the point ${\rm P}$ to the points 
between ${\rm P}$ and ${\rm P}'$ along $\cS$. 
Therefore, the $N$-th power of $D_q$ brings $\theta$ 
to the same position $\theta$ in $\cS$, \ie, 
$D_q^N : (x, \theta)\,\hookrightarrow \,(x, \theta+2\pi)$. 
However, upon taking the external space $\R$ into accout, 
the operator $D_q^N/[N]!$ becomes equivalent to 
the differential operator $\del_x$, the generator of the 
infinitesimal displacement along $\R$.   
Namely, for an infinitesimal number $\epsilon \in\R$, 
we find 
$
\frac{D_q^N}{[N]!} =\del_x : 
(x, \theta) \, 
    \mapsto \,(x+\epsilon, \theta)$.  
One can, therefore,  identify
$(x,\theta+2\pi)$ with  $(x+\epsilon, \theta)$.  
This identification indicates that 
the internal space $\cS$ connects two real points  
$x$ and $x+\epsilon$ which are separated 
by infinitesimal distance along $\R$. 
In this sense, the fuzzy space $\cS$ can be called  
the {\em infinitesimal structure}.  
In order to make such a structure more concrete 
and finally obtain the quantum real line $\vR$, 
some manipulations are needed.


The procedure for deriving $\vR$ from $\qR$ 
follows that given in the preceding subsection 
where $\R_{\rm D}$ was obtained from $\RRr$. 
We first reduce $\qR$ to a discrete $N$-point space $\qRN$ 
and, then,  $\qRN$ is modified 
to the final quantum real line $\vR$.  
To start with the first step, 
notice that $\cS$ is continous space. 
Actually, the variable $\xi$ parameterizing $\cS$ is continuous. 
However,  once a point $\xi_0\in \cS$ is chosen, 
a subalgebra $\cA_{\rm int}(\Imm(\xi_0))$ 
is obtained owing to the $q$-differential, \ie, 
difference structure. 
Here $\Imm(\xi_0)=\{ \psi(\xi_r)\, \vert \, 
      \xi_r=\xi_0 e^{2\pi i\frac{r}{N}}, 
      \, r=0, 1, \cdots, N-1\}$.  
One then finds the discrete space 
$\cS_N(\xi_0)=\{ \xi_r\,\vert \, r=0,1,\cdots, N-1\}$ 
as the base space of $\cA_{\rm int}(\Imm(\xi_0);D_q)$.  
If another point $\xi_0'$ is chosen, 
another subalgebra $\cA_{\rm int}(\Imm(\xi_0');D_q)$ 
and its base pace $\cS_N(\xi_0')$ are obtained. 
The continuous fuzzy space $\cS$ is the union 
of all the discrete spaces as 
$\cS=\bigcup_{0\leq \xi_0< q} \cS_N(\xi_0)$.  
Therefore, the base space $\qR$ is reduced to 
the discrete space 
\be
\qRN=\R\times \cS_N. 
\ee
The important fact to be noticed here is that, 
$\cS_N$ is also fuzzy space, \ie,   
each point in $\cS_N$ is not a stable and localized point. 
For example, the $r$-th point is  not located at 
the position $\xi_r$ but fluctuating around $\xi_r$.   

Let us go ahead to the next step where  $\qR_N$ 
is modified into the final quantum line $\vR$. 
This is performed by changing the internal space 
$\cS_N$ into another discrete space 
$\Sigma_{\check{\epsilon}}$. 
In the quantum space $\qR_N$, 
the internal space $\cS_N$ is attached 
on every point of $\R$ as the extra dimension. 
On the other hand, 
upon considering $\vR$ to be an extension of $\R$,  
an infinite number of the internal spaces $\sg$ are 
embedded into the blanks $\vR\setminus\R$. 
The space $\sg$ is defined by the mapping 
$\check{\pi} : \cS_N\,\rightarrow\,\sg$ such as  
\be
  \cS_N \ni \xi_r \quad  
     \longmapsto \quad \zeta_r = r{\check{\epsilon}} \in \sg,  
        \qquad  r=0,\,1,\,\cdots, N-1, 
    \label{SN}     
\ee
with some \lq\lq number" ${\check{\epsilon}}$. 
By writing the algebra of functions on $\sg$ as 
$\Imc$ and the difference operator with respect to $\zeta$ 
as $\Delta_{\check{\epsilon}}$, 
we have the algebra $\cA_{\rm int}(\Imc;
      \Delta_{\check{\epsilon}})$. 
The difference operator 
$\Delta_{\check{\epsilon}}$ is given  by 
\be
\Delta_{\check{\epsilon}} \, \phi(\zeta_r) = 
       \frac{\phi(\zeta_{r+1}) 
           - \phi(\zeta_r)}{{\check{\epsilon}}}, 
           \qquad \phi(\zeta) \in \Imc, 
\ee        
The point we should stress here is that, 
since $\sg$ is to be embedded into $\vR\setminus\R$, 
the number $\ce$ cannot be 
a real number but $\ce\in\vR\setminus\R$.  

It is the time to give the explicit definition 
of the quantum real line $\vR$.    
First, we put $\zeta_0$ on each real number $x\in \R$ 
and $\zeta_N:=\zeta_{N-1}+\check{\epsilon}$ 
on the real number $x+\epsilon$ 
with an infinitesimal number $\epsilon$. 
In this way, an internal  space $\sg$ is embedded 
into the \lq\lq crack" between $x$ and $x+ \epsilon$ 
for every real number $x\in\R$, 
and we define the space $\cM(x)$ by 
\be
\cM(x) = \{ \check{x}_r\, \vert \, 
     \check{x}_r = x + \zeta_r = x+r\check{\epsilon}, 
       \,\, r=0, 1,\cdots, N-1\}. 
\label{defM}    
\ee
We have finally reached the stage 
to propose $\vR$ through $\cM(x)$,  \\[.3cm]
{\bf Proposition 2} : Quantum Real Line $\vR$ 
                                                  \\[.1cm] 
{\em The qantum real line $\vR$ 
for the case with the deformation parameter $q$   
at the $N$-th root of unity  
is given by the union of $\cM(x)$ as},  
\be
\vR  = \bigcup_{x\in\R} \,\, \cM(x). 
\label{Rspace}      
\ee
\rule{13.55cm}{0pt} \rule{2mm}{2mm} \\[.13cm] 
The quantum real line $\vR$ is shown in Fig.2. 
\bee
\overbrace{ \rule{4cm}{0pt}   }^{\cM_x} 
                  \rule{2.75cm}{0pt} \nonumber \\[-.2cm]
\begin{array}{ccccccc}
\cdots & \circ & \bullet & \circ & \cdots & 
     \cdots & \circ 
                                             \\
       &       &  x-\epsilon &       &        &   &      
\end{array}
\begin{array}{ccccccc}
\bullet & \circ & \circ & \cdots & \cdots & 
     \circ & \bullet 
                                             \\
  x      & \check{x}_1 & \check{x}_2   &       &    
      &    &  x+\epsilon    
\end{array}
\begin{array}{cc}
\circ  &  \cdots      \\   & 
\end{array}&&   \nonumber
\eee
\begin{center}
Fig.2 \quad  The quantum real line  $\vR$
\label{fig1}      
\end{center} 
where the dots $\bullet$ stand for the 
\lq\lq standard" real numbers, 
while the circles $\circ$ represent numbers in 
$\vR\setminus \R$. 

Some remarks on $\vR$ are in order. 
We have called the internal space $\cS$ in $\qR$ 
the infinitesimal structure. 
In $\vR$, the infinitesimal structure lying 
between $x$ and $x+\epsilon$, $x,\epsilon\in\R$, 
corresponds to the spaces $\cM(x)\setminus\{\check{x}_0\}$. 
Thus, the infinitesimal structure spreads the interval 
between two real numbers $x$ and $x+\epsilon$.  
As we stated above, the number $\ce$ is an element of 
$\vR\setminus\R$. 
Since, $\epsilon$ is represented by $\ce$ as $\epsilon=N\ce$, 
one should not consider $\epsilon$ as an infinitesimal 
number in $\R$ but an element of $\vR$. 
In other words, 
from the viewpoints of the extended real numbers $\vR$, 
$\epsilon$ is no longer an infinitesimal but 
a measurable number. 
One regards $\epsilon$ as an infinitesimal number 
only when it is observed from the viewpoints of $\R$.

As can be seen from Fig.1 and Fig.2, 
both quantum real lines $\R_{\rm D}$ and $\vR$ are 
discrete spaces. 
However, there are essential differences between the two. 
In the final section, we will summarize these quantum lines 
by stressing the differences.

\section{Concluding Remarks}

In this paper, we have proposed possible deformations  
of the real numbers $\R$ by taking quantum effects into account, 
\ie, quantum real lines. 
Our discusstions started from deformation of 
the Heisenberg algebra $\cL$ by introducing an 
internal space $\cT$. 
In order to derive nontrivial quantum effects from 
the internal sector,   
we further introduce the Moyal product $\ag$ 
as the multiplication of the functional algebra on $\cT$. 
As the result, we have obtained the $q$-deformed 
Heisenberg algebra $\cLLq$ with the defining 
commutation relation (\ref{newcr}). 
We have next derived the algebras $\Dq$ 
of $q$-differentiable functions for the cases 
(I) where $q$ is generic, \ie,  $q=e^{i\gamma}$ 
with $\gamma$ being irratinal or pure imaginaty, 
(I$\!$I) where $q$ is a root of unity, \ie, 
$q=e^{2\pi i/N}$ with an integer $N$.  
From these algebras,  the geometrical structures 
of the base spaces have been deduced.  
Through these base spaces, 
we have proposd $q$-deformed real numbers, 
\ie, quantum real lines.  

Let us summarize our derivations of the 
quantum real lines. 
In the case (I), 
the algebra $\Dq(\Im^q;D_q)$ is 
the space of functions on which the action of the 
$q$-differential opeeator $D_q$ is defined. 
Here, the base space was introduced as  
$\RRr=\{\xx\, \vert \, \xx = x_+ e^{i\theta}\}$.  
We have shown  that noncommutative natures of the algebra, 
e.g., eq.(\ref{NC}), arrise from the $q$-differential structure. 
In order to show the geometrical structures of the base space 
in explicit, 
the algebra $\Dq(\Im^q;D_q)$ on $\RRr$ was modified 
to the algebra $\Dq(\Img;\Delta_\gamma)$, 
where the base space was modified to $\RR$ as well. 
Upon parameterizing $\RR$  by the variable $\zx$ 
given in (\ref{defzx}),  
$\RR$ is still a continuous space. 
However, we found an infinite number of subalgebras 
in $\Dq(\Img;\Delta_\gamma)$. 
Actually, by fixing a point $\zx_0$ as an origin,  
the subspace 
$\Im_{\rm D}(\zx_0) =\{ f(\zx_n)\, \vert \, 
   \zx_n=\zx_0 + n\gamma, \, n \in \Z\}$ 
is extracted from $\Img$. 
Since two subalgebras $\Im_{\rm D}(\zx_0)$ 
and $\Im_{\rm D}(\zx_0')$, $\zx_0'\neq \zx_0+n\gamma$ 
are equivalent,  
we have divided $\Dq(\Img;\Delta_\gamma)$ by 
the equivalence and finally reached the algebra 
$\Dq(\Im_{\rm D}:\Delta_\gamma)$.  
Hence, the base space of $\Dq(\Im_{\rm D}:\Delta_\gamma)$ 
is the discrete space $\R_{\rm D}$ shown in Fig.1.  
We have proposed $\R_{\rm D}$ 
as the $q$-deformed real line 
in the case with generic $q$.  
The process from $\RRr$ to $\R_{\rm D}$ is 
\be
\begin{array}{ccccc}
\RRr & \longrightarrow & \RR & \searrow & \R_{\rm D} 
                 \\[-.2cm]
\upin &                & \upin &               & \upin      
                \\[-.2cm]
\xx   &                & \zx   &               & \zx_n \,\, 
                                                 (n\in \Z)
\end{array}
\label{sum1}    
\ee
where $\searrow$ stands for the reduction.

Let us nnext summarize the case where 
$q$ is the $N$-th root of unity. 
We have started with the algebra $\Im^q_N$ 
of functions on the base space 
$\RRn=\{\xxx\,\vert\, \xxx=y\xi,\, \xi=e^{i\theta}\}$. 
When the differential structure to be introduced 
in $\Im^q_N$ was considered, 
we found the identity $D_q^N\equiv 0$ on $\Im_N^q$.   
On the contrary,  we further found that the operator 
$D_q^N/[N]!$ acts on $\Im^q_N$ as 
$\frac{D_q^N}{[N]!}y^{kN+r}=k y^{(k-1)N+r}$.  
From these observations, we have seen that 
the algebra $\Dq_N(\Im^q_N;D_q)$ on $\RRn$ 
decomposes as $\Dq_N(\Im^q_N;D_q) \rightarrow \DqN=
   A(\Im;\del_x)\otimes \cA_{\rm int}(\Im_{\rm int};D_q)$ 
where $A(\Im;\del_x)$ is the standard algebra of 
differentiable functions 
on $\R=\{ x\,\vert\,x=y^N\}$. 
On the other hand, $\cA_{\rm int}(\Im_{\rm int};D_q)$ is 
the algebra of $q$-differentiable functions 
on the internal space 
$\cS=\{\xi\,\vert\, \xi=e^{i\theta}, \, 0\leq \theta < 2\pi\}$. 
The functional algebra $\Imm$ on $\cS$ has the nonstandard 
property that the basis $\varpi=\{\xi^n\,\vert\, n\in \cI\}$ 
of $\Imm$ is of finite dimention, 
\ie, ${\rm dim}.\varpi=N$. 
This property affects the structure of 
$\cS$. 
Indeed, we have seen that $\cS$ is a fuzzy space. 
An important fact is that the internal space $\cS$ is 
attached on every point of the real line $\R$ 
and connects two real numbers $x$ and $x+\epsilon$ 
with an infinitesimal $\epsilon$.  
Thus, we have concluded that 
the base space $\RRn$ should be transformed to 
$\qR$ of the algebra $\DqN$ and $\qR$ is the product 
space as $\qR=\R\times \cS$.  
To reach the finial quantum real line, 
two steps were needed for the internal sector 
$\cA_{\rm int}(\Imm;D_q)$. 
The first was the reduction to 
$\cA_{\rm int}(\Im(\xi_0);D_q)$ composed of 
functions on the discrete space 
$\cS_N=\{\xi_r\,\vert\, \xi_r=\xi_0q^r,\, r=0,1,\cdots,N-1\}$, 
and we obtained the base space $\qR_N=\R\times \cS_N$.  
The next step modified $\cA_{\rm int}(\Im(\xi_0);D_q)$ to 
the algebra 
$\cA_{\rm D}(\check{\Im};\Delta_{\check{\epsilon}})$ 
where $\Imc$ was the algebra of functions on the space 
$\sg=\{\zeta_r\,\vert\, r=0,1,\cdots,N-1\}$. 
The space $\sg$ was introduced instead of $\cS_N$ so that 
we can extend $\R$  to $\vR$. 
The base space $\vR$ is just 
the quantum real lne we have proposed . 
To define $\vR$ explicitly, the space $\cM(x)$ was 
introduced at ech point $x\in\R$ as 
$\cM(x)=\{\check{x}_r\,\vert\, 
        \check{x}_r=\check{x}_0+r\ce, \, \check{x}_0=x,\,
         r=0,1,\cdots, N-1\}$. 
Finally, the quantum real line in the case 
with $q$ at the $N$-th root of unity 
has been defined by $\vR=\bigcup_{x\in\R} \cM(x)$. 
Thus, we have found the infinitesimal structure 
represented by $\cM(x)$ 
between two points $x,\,x+\epsilon \in\R$ 
with an infinitsimal number $\epsilon$. 
In summary, the extension $\vR$ of real numbers $\R$ 
has been performed by the procedure as; 
\be
\begin{array}{cclclcl}
\RRn & \longrightarrow & \qR\,\left(\R\times \cS\right) 
    & \searrow  
    & \qR_N\,\left( \R\times \cS_N \right) 
    & \longrightarrow 
    & \vR \, 
      ( \bigcup_{x\in\R} \cM(x)\,) \\[-.2cm] 
\upin &  {}  & \quad \upin & {}   & \,\, \upin  & {} 
                   & \quad \upin     \\[-.2cm]
\xxx &  {}  & \, (x,\xi)   & {} & \, (x,\xi_r) & {} 
              & \quad \check{x}_r \,\, 
\end{array}
\label{sum2}   
\ee

Let us give some remarks upon 
natures of the quantum real numbers $\R_{\rm D}$ and $\vR$, 
and we would like to stress the differences between the two.  
Although both have similar discrete structures,  
there exist some essential differences. 
The fact to be stressed first is that 
the quantum line $\R_{\rm D}$ is, in general,  
composed of standard complex numbers, \ie, $\zx_n\in\C$ 
and points $\zx_n$ are localized objects.  
The quantum effect appears in $\R_{\rm D}$ as the 
blank between two adjacent points. 
We have discussed in Section 4.1 
the physical picture of the interval, \ie, 
it is just the {\rm inimal length} in quantum line. 
In other words, we always have the uncertainty 
$\Delta\chi\sim\gamma$ in measurement of position. 

On the other hand, 
each point $\check{x}_r\in \vR$ 
has fuzzy structure. 
Furthermore,  $\check{x}_r$ cannot be represented 
in terms of the standard complex numbers, 
especially the interval $\ce$ between $\check{x}_r$ and 
$\check{x}_{r+1}$ is the case. 
We have stressed that $\vR$ is an extended object 
of the real number $\R$. 
However, in $\vR$, 
one can find $\R$ itself as the subset of $\vR$. 
Actually, the dots $\bullet$ in Fig.2 compose $\R$. 
It is quite important to notice that 
the number $\epsilon$, the interval beteween 
two adjacent $\bullet$, is an 
infinitesimally small number as long as  we  
observe it within the viewpoints of $\R$. 
On the contrary, once we observe $\epsilon$ 
from the standpoint of $\vR$, it is not an infinitesimal 
but an measurable number 
in terms of $\ce$ as $\epsilon =N\ce$. 

Another characteristic feature of $\vR$ to be emphasized again 
is the existence of the internal space 
$\cM(x)\setminus\{\check{x}_0\}$ 
at each pint $\check{x}_0 \equiv x\in \R$. 
In Fig.2, the numbers in the internal spaces 
are expressed by $\circ$.  
We have called the internal space 
the {\em infinitesimal structure} in the sense that 
$\cM(x)\setminus \{\check{x}_0\}$ lives 
between the two real numbers $x$ and $x+\epsilon$. 
Thus, the quantum effects appearing in $\vR$ is different 
from those in $\R_{\rm D}$.  
The discreteness yielding the noncommutativity 
and the minimal length is considered as 
the only quantum effect in $\R_{\rm D}$, 
while, in $\vR$, 
the appearance of the infinitesimal structure is the 
characteristic quantum effect. 

Let us find another difference between 
$\R_{\rm D}$ and $\vR$ by looking at them at a long-scale. 
As for the quantum line $\R_{\rm D}$, 
we take the long-scale limit through $q \rightarrow1$, \ie,  
$\gamma\rightarrow0$.   
Then, the interval between two points becomes 
infinitesimally small and, therefore, 
the quantum line $\R_{\rm D}$ reduces to $\R$. 
Namely, at the long-scale, 
the discreteness of $\R_{\rm D}$ is not visible 
and $\R_{\rm D}$ looks continuous. 
On the other hand, for the line $\vR$, 
one obtains the case with $q=1$ by setting $N=1$. 
Then, $\cM(x)$ bocomes the one-point space 
$\cM(x)\vert_{N=1} =\{\check{x}_0=x\in\R \}$,  
\ie, the internal space vanishes and  $\vR=\R$ 
with $\ce=\epsilon$ being an infinitesimal.  
The above situation is quite trivial and not beyond 
our expectation. 
However, in $\vR$, there is another interesting case 
where we take the limit 
$\epsilon\rightarrow0$ with keeping $N\neq 1$. 
This case corresponds to the observation of $\vR$ 
at the long-scale such that the interval $\epsilon$ is 
infinitesimally small and invisible. 
In this limit, each internal space $\cM(x)$ shrinks to 
a point $x\in\R$ and  
we find the real line in which 
every point $x$ has the internal structure.   

Let us end this paper by addressing an interesting problem 
to be studied further. 
It is the application of the quantum real line $\vR$ 
with $N=2$ to supersymmetric model.  
Actually, the author has shown the equivalence 
between the quantum universal emvelopping algebra 
$\qslc$ with $q$ at the 2nd root of unity and 
the supersymmetric algebra ${\rm Osp}(2\vert1)$ \cite{TS2}. 
In the case with $N=2$, the space $\cM(x)$ is the two-point 
space such as $\cM(x)=\{ \bullet,\,\circ\}$. 
It is natural to expect that one 
of them corresponds to bosonic space and 
the other to fermionic space. 
The investigation will appear elsewhere.

\vspace{1cm}


\noindent
{\bf Acknowledgments} \\[.15cm]
The author would like to thank Dr. T. Koyama 
for valuable comments.



\begin{thebibliography}{99}

\bibitem{Co} ADConnes, 
     \lq\lq {Noncommutative Geometry}", 
     Academic Press (1994). 

\bibitem{Ma} Yu. I. Manin, 
   \lq\lq{Topics in Nongommutative Geometry}", 
   Princeton University Press  (1991).  

\bibitem{TS1}  T. Suzuki,  
  \lq\lq{Doformations of Poisson algebra 
  in terms of star products and $q$-deformed mechanics}", 
    J. Math. Phys. {\bf 34} (1993) 3453. 


\bibitem{TS2} T. Suzuki,  
   \lq\lq{More on $U_q({\rm su}(1,1))$ 
   with $q$ a root of unity}", 
  J. Math. Phys. {\bf 35} (1994) 6857. 

 
\end{thebibliography}
\end{document}